\documentclass[refereed]{aa} 
\usepackage{times,amssymb,amsmath} 
\usepackage{epsfig} 

\title{A Lyman-$\alpha$ blob in the GOODS
South field: evidence for cold accretion onto a dark matter halo
\thanks{Based on observations carried out at the European Southern
Observatory (ESO) under prog. ID No. 70.A-0447 and 274.A-5029.}}

\author{K.~K. Nilsson\inst{1,2}
        \and J.~P.~U. Fynbo\inst{2}
        \and P. M\o ller\inst{1}
        \and J. Sommer-Larsen\inst{2}
        \and C. Ledoux\inst{3}
}

\institute{
   European Southern Observatory, Karl-Schwarzschild-Stra\ss e 2, 85748
   Garching bei M\"unchen, Germany
\and
   DARK Cosmology Centre, Niels Bohr Institute, University of Copenhagen, 
   Juliane Maries Vej 30, 2100 Copenhagen $\O$, Denmark
\and
   European Southern Observatory, Alonso de C\'ordova 3107, Casilla
   19001, Vitacura, Santiago 19, Chile
}
\offprints{knilsson@eso.org}
\mail{knilsson@eso.org}
\date{Received date / Accepted date}
\begin{document} 
\titlerunning{A Ly$\alpha$ blob in the GOODS South field}

\abstract{
We report on the discovery of a $z = 3.16$ Lyman-$\alpha$ emitting blob in the 
Great Observatories Origins Deep Survey (GOODS) South field. 
The discovery was made with the VLT, through
 narrow-band imaging.
 The blob has a total Ly$\alpha$ luminosity of 
$\sim10^{43}$~erg~s$^{-1}$ and a diameter larger than 60~kpc. 
The available multi-wavelength data in the GOODS field consists of 
13 bands from X-rays (Chandra) to infrared (Spitzer). Unlike 
other known Ly$\alpha$ blobs, this blob shows no obvious 
continuum counter-parts in any of the broad-bands. In particular, 
no optical counter-parts are found in deep HST/ACS imaging. For 
previously published blobs, AGN (Active Galactic Nuclei) or ``superwind'' 
models have been found to provide the best
match to the data. We here argue that the most probable origin of the
extended Ly$\alpha$ emission from this blob is cold 
accretion onto a dark matter halo.  
}

\maketitle

\section{Introduction}
Narrow-band surveys for Lyman-$\alpha$ (Ly$\alpha$) emitting galaxies at high
redshift have recently revealed a number of luminous (up to $5 \cdot 10^{43}$ 
erg s$^{-1}$), very extended (from a few times ten kpc to more than 150 kpc)
Ly$\alpha$-emitting objects, so-called Ly$\alpha$ ``blobs'' (Fynbo et al. 1999;
Keel et al. 1999; Steidel et al. 2000; Francis et al.  2001; 
Matsuda et al. 2004; Palunas et al.
2004; Dey et al. 2005; Villar-Martin et al. 2005). 
At least three mechanisms have been suggested as energy sources for 
Ly$\alpha$ blobs. These are: \emph{i)} hidden QSOs (Haiman \& Rees 2001; 
Weidinger et al. 2004, 2005), \emph{ii)} star formation and superwinds from
(possibly obscured) starburst galaxies (Taniguchi et al. 2001; Ohyama et al.
2003; Mori et al. 2004; Wilman et al. 2005), and \emph{iii)} so-called cold
accretion (Haiman, Spaans \& Quataert 2000; Fardal et al. 2001; Keres et al.
2004; Maller \& Bullock 2004; Birnboim \& Dekel 2003; Sommer-Larsen 2005; 
Dijkstra et al. 2006(a,b); Dekel \& Birnboim 2006).
Cooling flows are phenomena observed in galaxy clusters for more than a decade
(Fabian 1994). These are explained by gas which is cooling
 much faster than the 
Hubble time through X-ray emission in the centres of the clusters. 
However, cooling
emission from a galaxy, or a group sized halo can be dominated by 
Ly$\alpha$ emission 
(e.g. Haiman, Spaans \& Quataert 2000; Dijkstra et al. 2006(a,b)).
In this \emph{Letter} we present the discovery of a
Ly$\alpha$ blob at redshift $z \approx 3.16$ located in the GOODS South
field, which we argue is the first piece of evidence for cold gas
accretion onto a dark matter halo.

Throughout this paper, we assume a cosmology with $H_0=72$ km s$^{-1}$
Mpc$^{-1}$, $\Omega _{\rm m}=0.3$ and $\Omega _\Lambda=0.7$. All magnitudes are
in the AB system.

\section{Observations and Data reduction}
\label{obs}
A 400$\times$400 arcsec$^2$ section, centred on R.A.~$= 03^h 32^m 21.8^s$, 
Dec~$= -27^{\circ} 45' 52''$ (J2000), of the GOODS South field
was observed 
with FORS1 on the VLT 8.2 m telescope Antu
during two visitor mode nights on December 1--3, 2002. 
A total of 16 dithered exposures were obtained over the two nights
for a combined exposure time of 30 ksec, all with the narrow band
filter OIII/3000+51 and using the standard resolution collimator
 (0.2$\times$0.2
arcsec$^2$ pixels). For this setup the central wavelength of the
filter is 5055 {\AA} with a FWHM of 59 {\AA}, corresponding to the
redshift range $z = 3.126$~--~$3.174$ for Ly$\alpha$.

The observing conditions were unstable during the two nights with the
seeing FWHM varying between 0\farcs66 and 1\farcs25 on the first night
and 1\farcs4 and 3\farcs3 on the second night.
The images were reduced (de-biased, and corrected for CCD pixel-to-pixel
variations using twilight flats) using standard techniques. 
The individual reduced images were combined using a modified version
of our code that optimizes the Signal-to-Noise (S/N) ratio for faint,
sky-dominated sources (see M{\o}ller \& Warren 1993, for details on this
code). The modification of the code was necessitated by the highly
variable seeing. The sky background was assumed to be constant. The 
FWHM of the PSF of the final combined narrow-band image is 0\farcs8.

For object detection, we used the software package SExtractor
(Bertin \& Arnouts 1996). 
A full description of our selection of Ly$\alpha$ emitters in the GOODS field
 will be given in a subsequent paper. In this {\it Letter} we 
discuss the nature of an extended, low surface brightness
blob with a centroid (of the Ly$\alpha$ emission) of 
R.A.~$ = 03^h 32^m 14.6^s$  and Dec~$ = -27^{\circ} 43' 02$\farcs$4$ (J2000) detected in the combined narrow-band image. 
  
Follow-up MOS spectroscopy was obtained in service mode using 
FORS1/VLT UT2 over the time period December 2004 -- February 2005. The 
total observing time was 6 hours. 
We used a 1\farcs4 slitlet and grism 600V resulting in a wavelength range of
4650 {\AA} to 7100 {\AA} and a spectral resolution FWHM of approximately 700. 
The seeing varied between 0\farcs77 and 1\farcs2 during the spectroscopic
observations.

The GOODS archival data used here and their detection limits are listed in
Table~\ref{mwtab}.

\begin{table} 
\begin{center}
\caption{Specifications of deep, multi-wavelength data available in the 
GOODS South field and the narrow-band image. 
The last column gives the 3$\sigma$ limit as detected in a $2''$ radius 
aperture and the narrow-band value gives 
the blob flux in this aperture. }
\vspace{-0.5cm}
\begin{tabular}{@{}lccccccc}
\hline
Filter/Channel & $\lambda_c$ & Filter & 3$\sigma$ limit ($2''$ aperture)\\
& & FWHM  & ($\mathrm{erg}$~$  \mathrm{cm}^{-2}$$\mathrm{s}^{-1}$$\mathrm{Hz}^{-1}$)\\
\hline
X-rays (\emph{Chandra})     &  4.15 keV    &  3.85 keV    & $9.90 \cdot 10^{-34}$   \\
U (\emph{ESO 2.2-m})        &  3630 \AA    &  760 \AA     & $8.62 \cdot 10^{-31}$  \\
B (\emph{HST})              &  4297 \AA    &  1038 \AA    & $9.25 \cdot 10^{-30}$ \\
Narrow (\emph{VLT})         &  5055 \AA    &  60 \AA      & $6.68 \cdot 10^{-30}$  \\
V (\emph{HST})              &  5907 \AA    &  2342 \AA    & $4.66 \cdot 10^{-30}$  \\
i (\emph{HST})              &  7764 \AA    &  1528 \AA    & $1.50 \cdot 10^{-29}$  \\
z (\emph{HST})              &  9445 \AA    &  1230 \AA    & $3.00 \cdot 10^{-29}$ \\
J (\emph{VLT})              &  1.25 $\mu$m &  0.6 $\mu$m  & $5.31 \cdot 10^{-30}$ \\
H (\emph{VLT})              &  1.65 $\mu$m &  0.6 $\mu$m  & $1.86 \cdot 10^{-29}$ \\
Ks (\emph{VLT})             &  2.16 $\mu$m &  0.6 $\mu$m  & $1.56 \cdot 10^{-29}$ \\
Ch1 (\emph{Spitzer/IRAC})   &  3.58 $\mu$m &  0.75 $\mu$m & $2.51 \cdot 10^{-31}$  \\
Ch2 (\emph{Spitzer/IRAC})   &  4.50 $\mu$m &  1.02 $\mu$m & $6.43 \cdot 10^{-32}$ \\
Ch3 (\emph{Spitzer/IRAC})   &  5.80 $\mu$m &  1.43 $\mu$m & $5.01 \cdot 10^{-29}$ \\
Ch4 (\emph{Spitzer/IRAC})   &  8.00 $\mu$m &  2.91 $\mu$m & $4.65 \cdot 10^{-30}$ \\
\hline
\label{mwtab} 
\end{tabular} 
\end{center} 
\vspace{-1.2cm}
\end{table}

\section{Results}\label{results}

\begin{figure*}[!ht] \begin{center} 
\epsfig{file=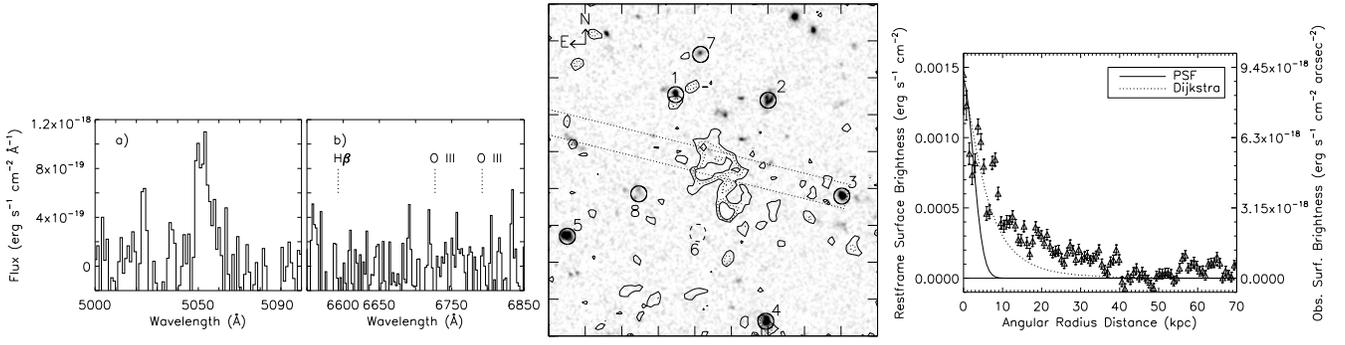,width=18cm,clip=}

\caption{\emph{Left}~\emph{a)} Flux calibrated spectrum of the blob emission 
line. The
line has the characteristic blue side absorption, indicating high redshift.
\emph{b)} The part of the spectrum (binned with a binsize equal to half 
the resolution ($1.1$~{\AA})) where H$\beta$ and [OIII] should have been observed if the emission line was [OII] at a redshift of $z \approx 0.36$. 
These lines are not observed and
therefore we conclude the observed line is due to Ly$\alpha$ at $z=3.16$. 
\emph{Middle} Contour-plot of narrow-band emission from the Ly$\alpha$ blob
overlaid the
HST V-band image. The narrow-band image has been continuum
subtracted by subtracting the re-binned, smoothed and scaled HST/V-band image. 
Contour levels are $2 \cdot 10^{-4}$, $4 \cdot 10^{-4}$ and $6
\cdot 10^{-4}$~erg~s$^{-1}$~cm$^{-2}$ in restframe flux (corresponding 
to $1.2 \cdot 10^{-18}$, $2.5 \cdot 10^{-18}$ and $3.7 \cdot 10^{-18}$ in 
observed flux). The image is $18'' \times 18''$ ($18''$ corresponds to a physical size of $\sim 133$~kpc).
Numbers refer to those used in section~\ref{results}. The dotted lines indicate
the slitlet position for our follow-up spectroscopy. 
%The slitlet also covered 
%galaxy \#~3, however the spectrum was too close to the end of the slitlet 
%to be observed.
\emph{Right} Plot of surface brightness as function of radius. 
The flux is the sky subtracted narrow-band flux.  
The PSF of the image is illustrated by the solid line, and the 
dotted line is the best fit model of Dijkstra et al. 2006. 
The deficit at $\sim 45$~kpc is due to the asymmetric appearance of 
the blob.}
\label{surfbright}
\label{spectrum} 
\label{contour}
\end{center} 
\vspace{-0.5cm}
\end{figure*}

The spectrum of the part of the Ly$\alpha$ blob covered by the
slitlet can be seen in the left-most panel of Fig.~\ref{spectrum}. 
The line has the asymmetric profile expected for a high redshift
Ly$\alpha$ emitter. 
We detect no other emission lines in the spectrum.
The most likely interloper is [OII] at redshift 0.36, but no emission
is observed in the spectrum where e.g. H$\beta$ or [OIII] are expected at this
redshift, see Fig.~\ref{spectrum}. This leads us to the
conclusion that we are observing a Ly$\alpha$-emitting object at $z = 3.157$. 
The observed FWHM velocity width of the emission line is $505$~km~s$^{-1}$. 
The instrument FWHM of the set-up is $290$~km~s$^{-1}$, hence 
the Ly$\alpha$ intrinsic velocity width is marginally resolved. The 
intrinsic width is less than $500$~km~s$^{-1}$. This is of the order or 
smaller than for other published blobs,
with velocity widths of $500 - 2000$~km~s$^{-1}$ (Keel et al. 1999;
Steidel et al. 2000; Francis et al. 2001; Ohyama et al. 2003; Bower et al 2004; 
Dey et al.  2005).

\begin{figure*}[!ht] \begin{center} 
\epsfig{file=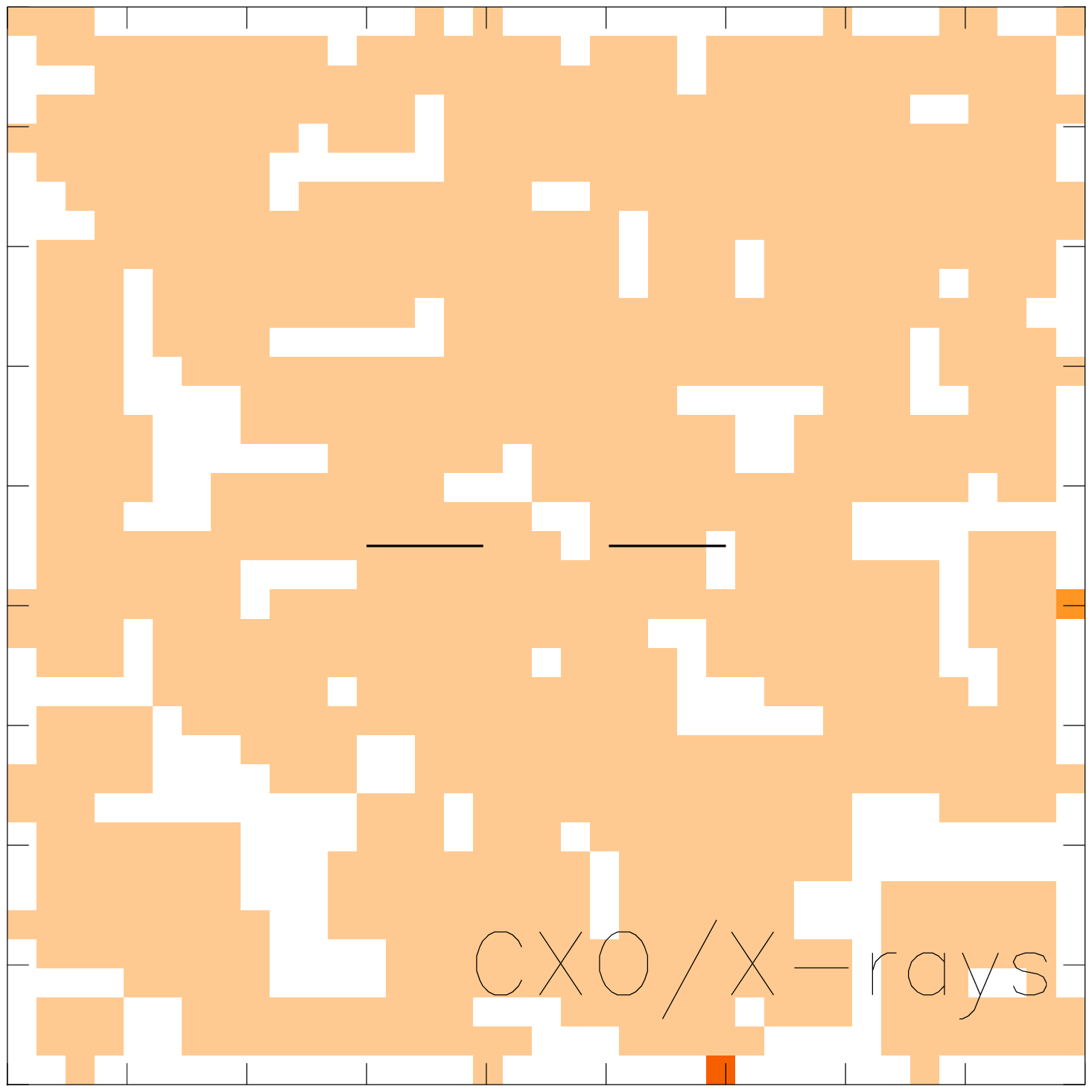,width=2.5cm,clip=}\epsfig{file=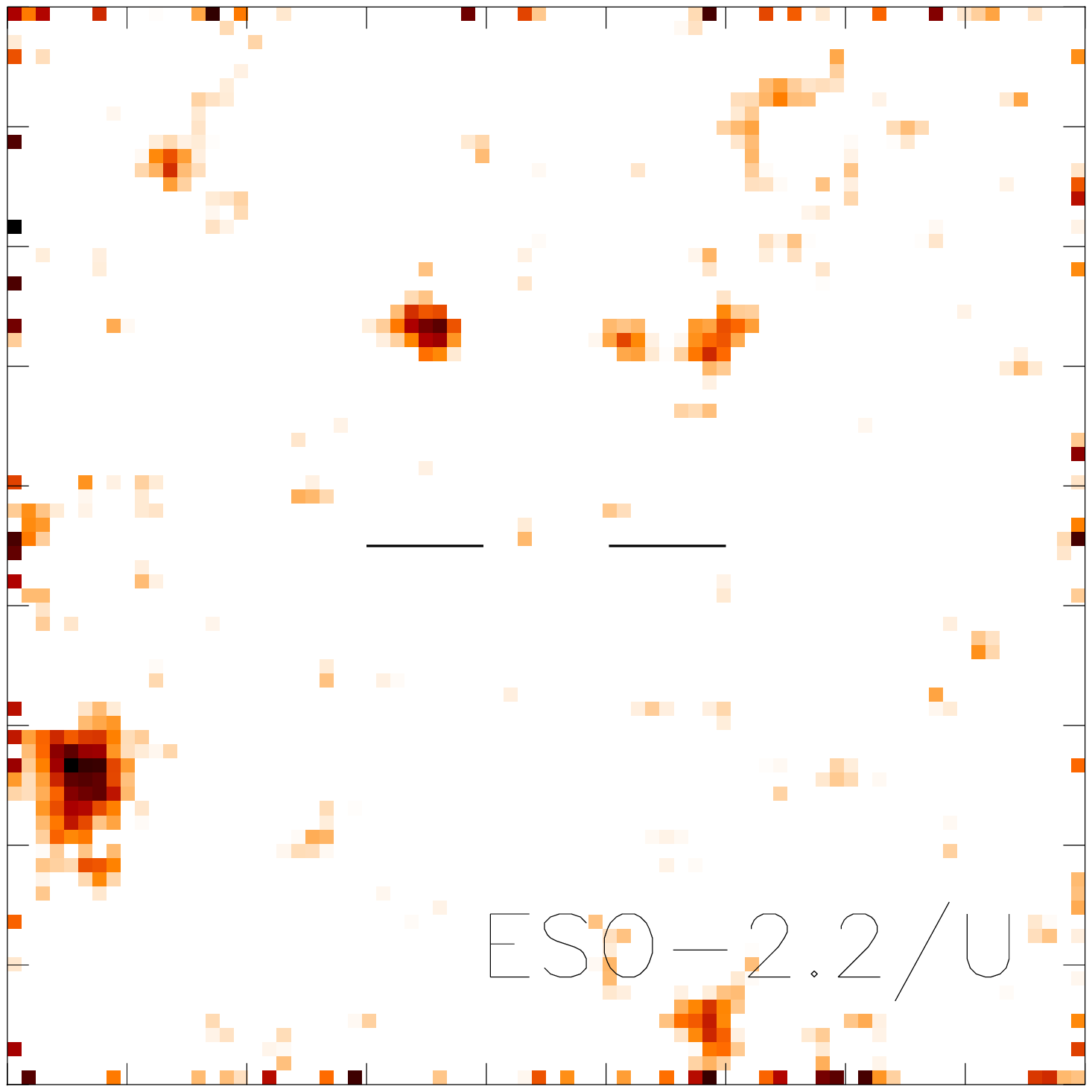,width=2.5cm,clip=}\epsfig{file=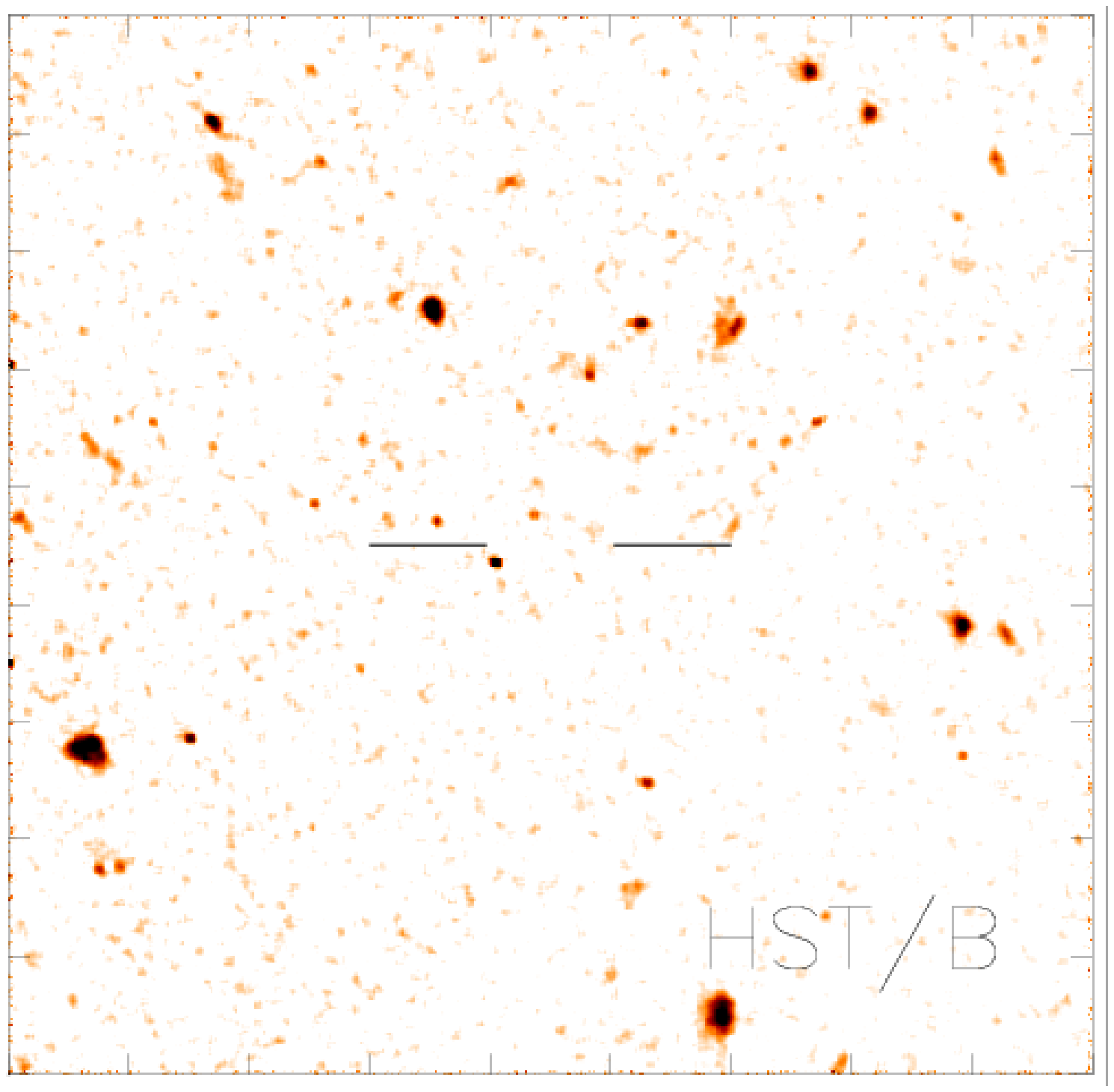,width=2.5cm,clip=}\epsfig{file=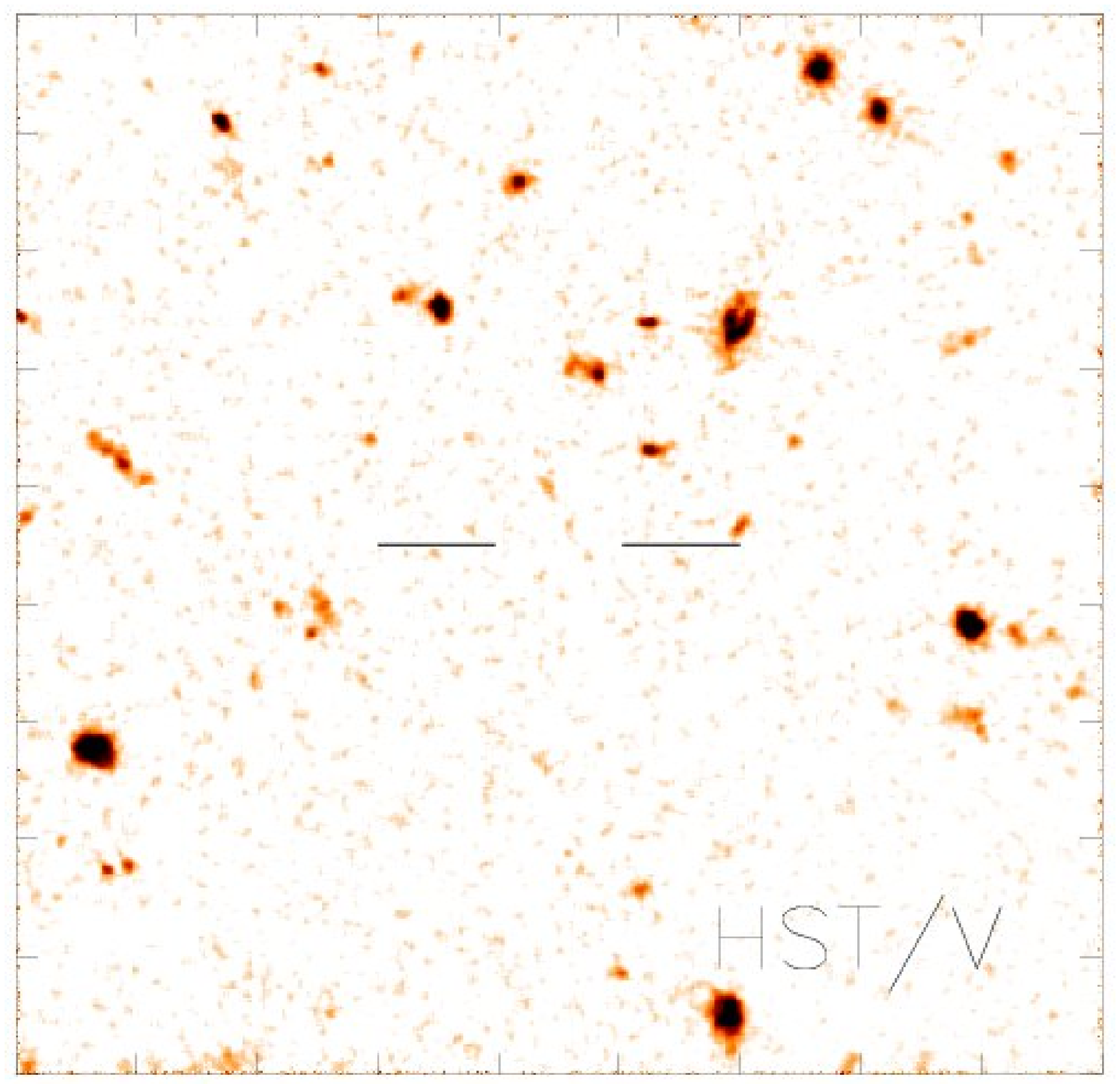,width=2.5cm,clip=}\epsfig{file=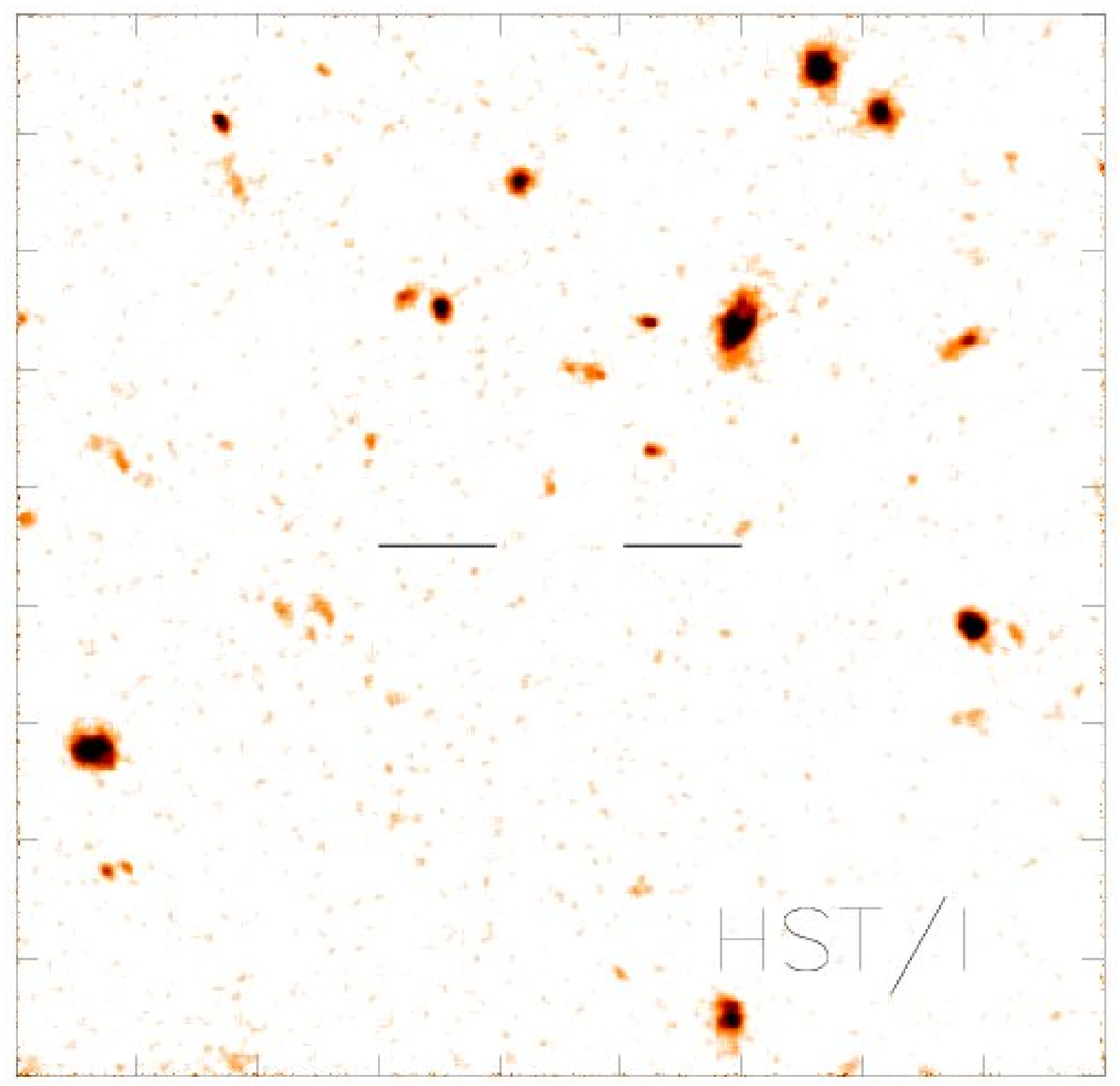,width=2.5cm,clip=}\epsfig{file=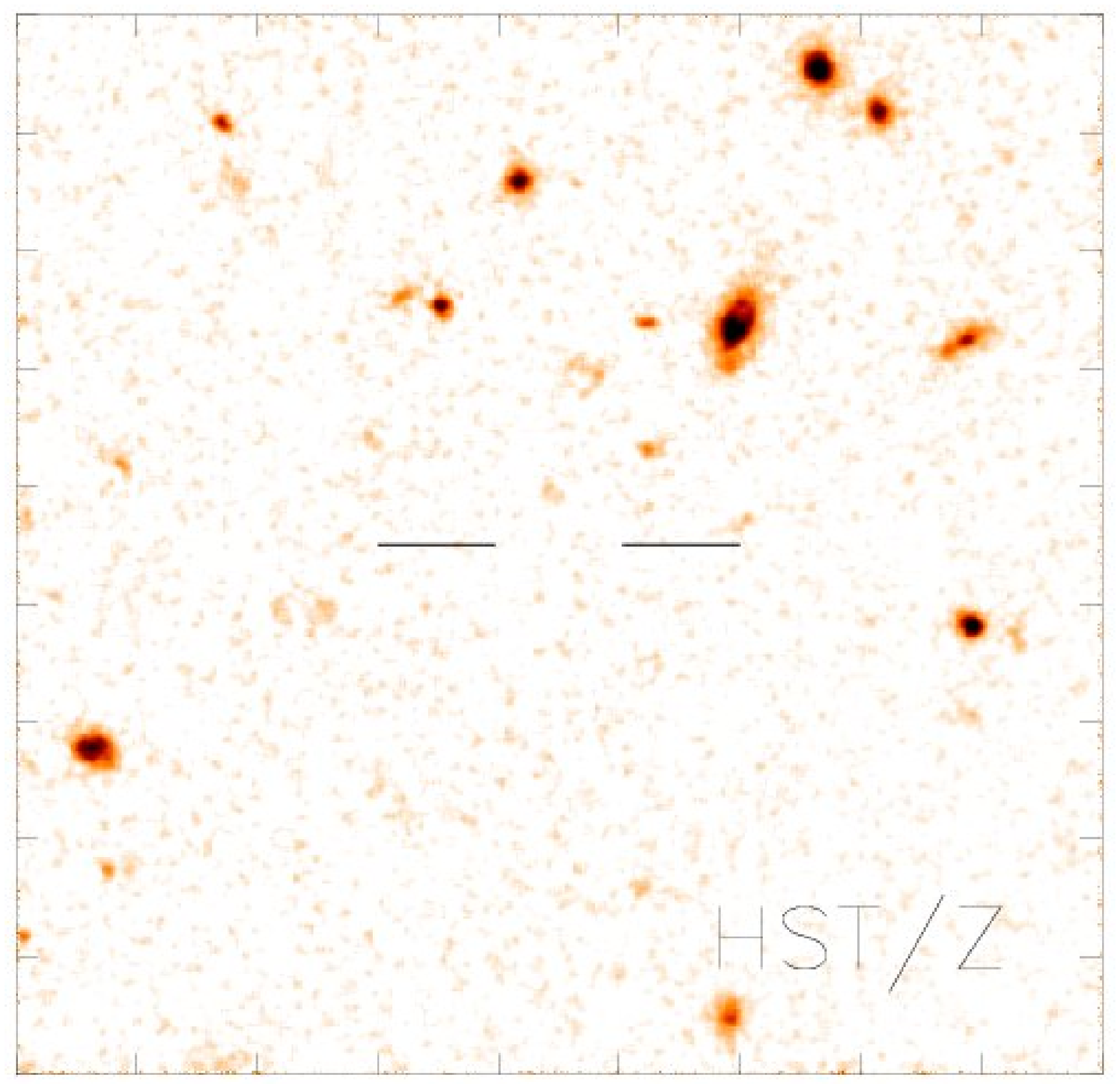,width=2.5cm,clip=}\epsfig{file=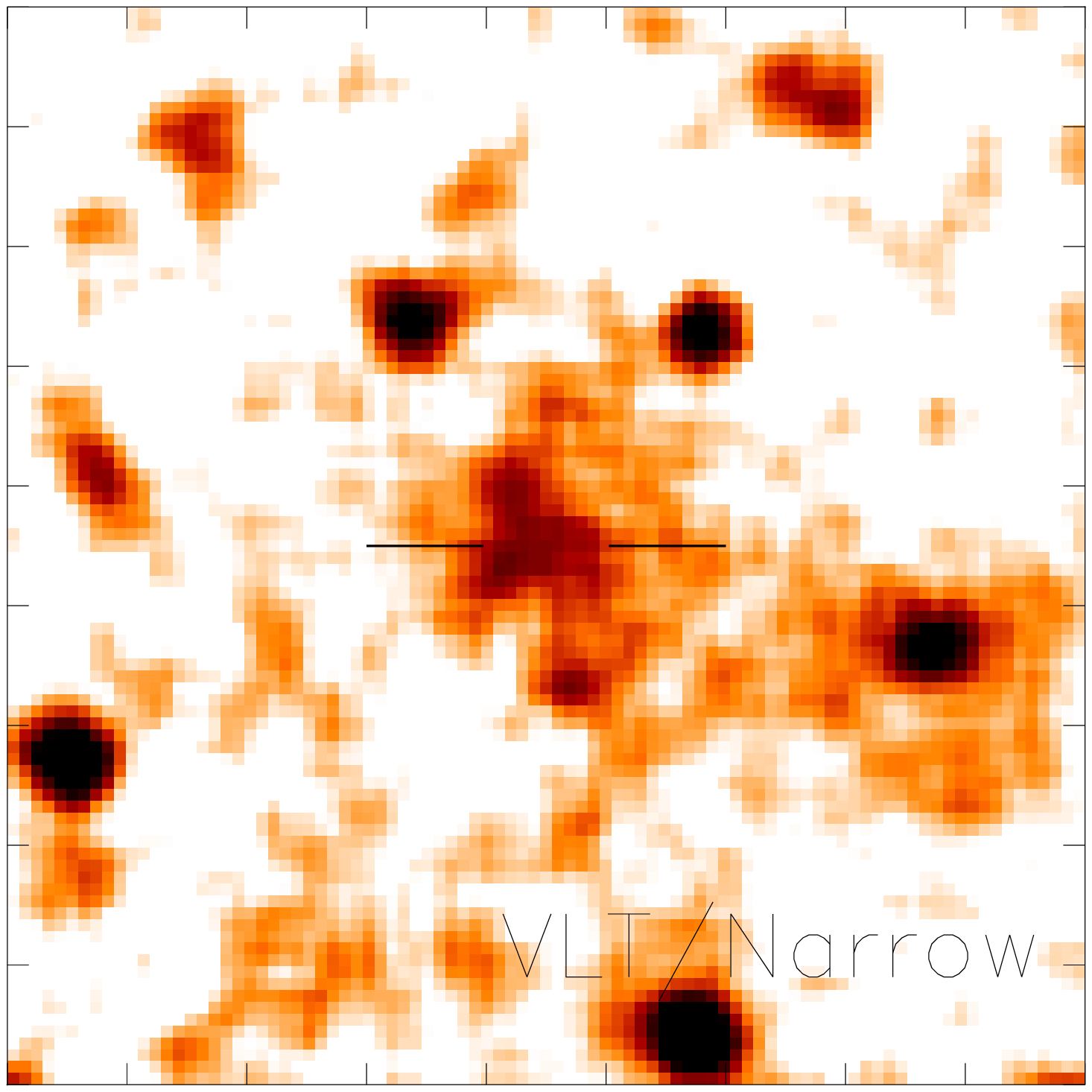,width=2.5cm,clip=}

\epsfig{file=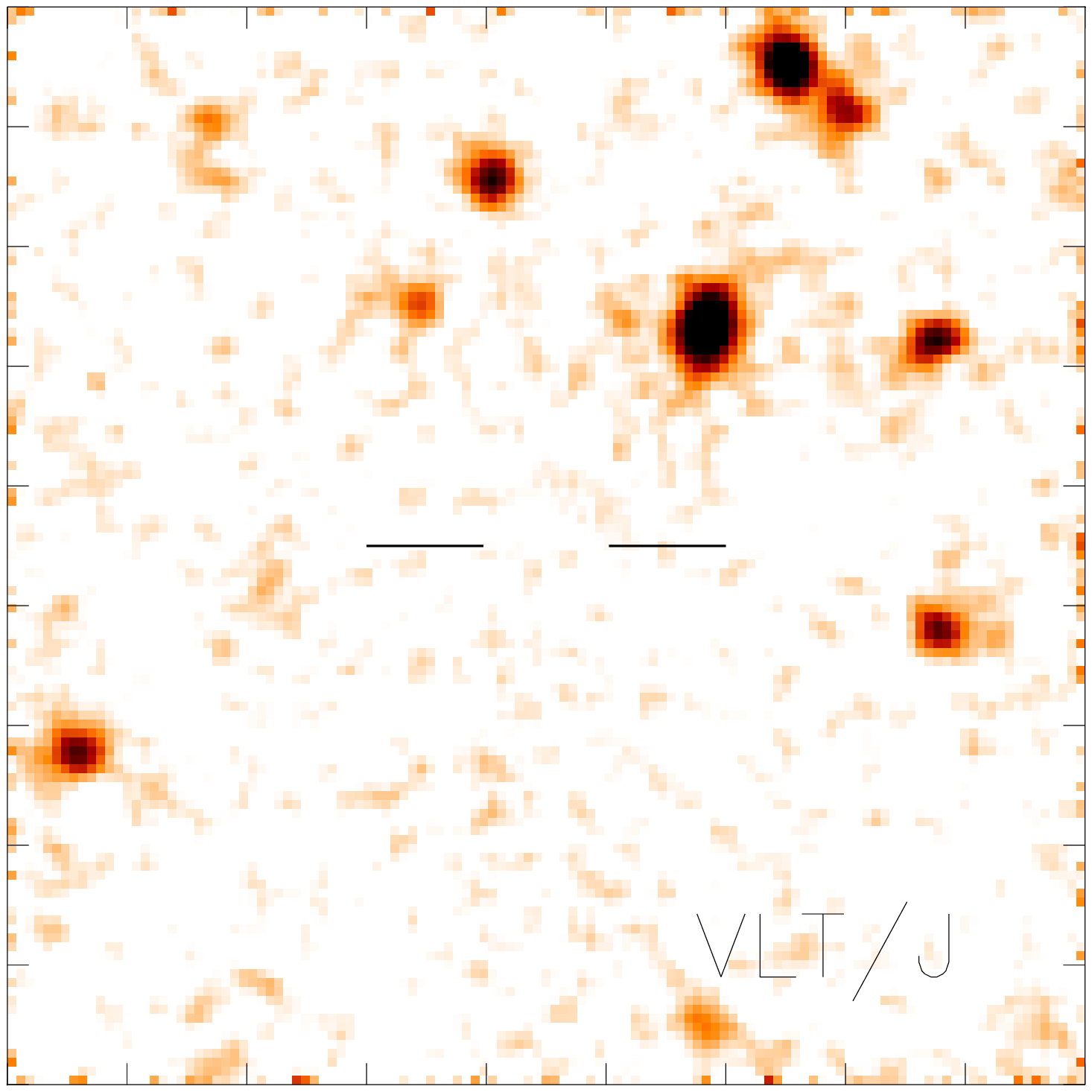,width=2.5cm,clip=}\epsfig{file=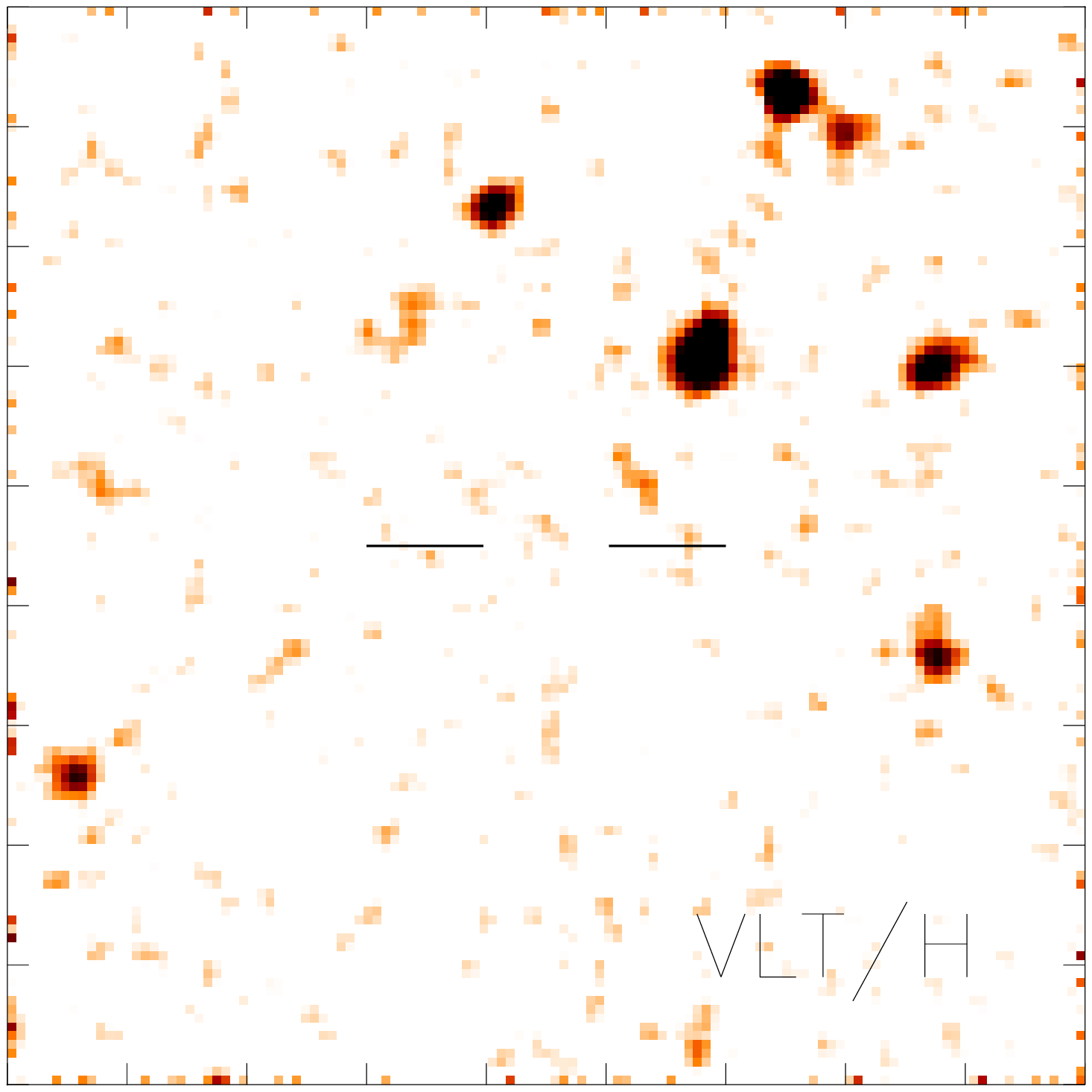,width=2.5cm,clip=}\epsfig{file=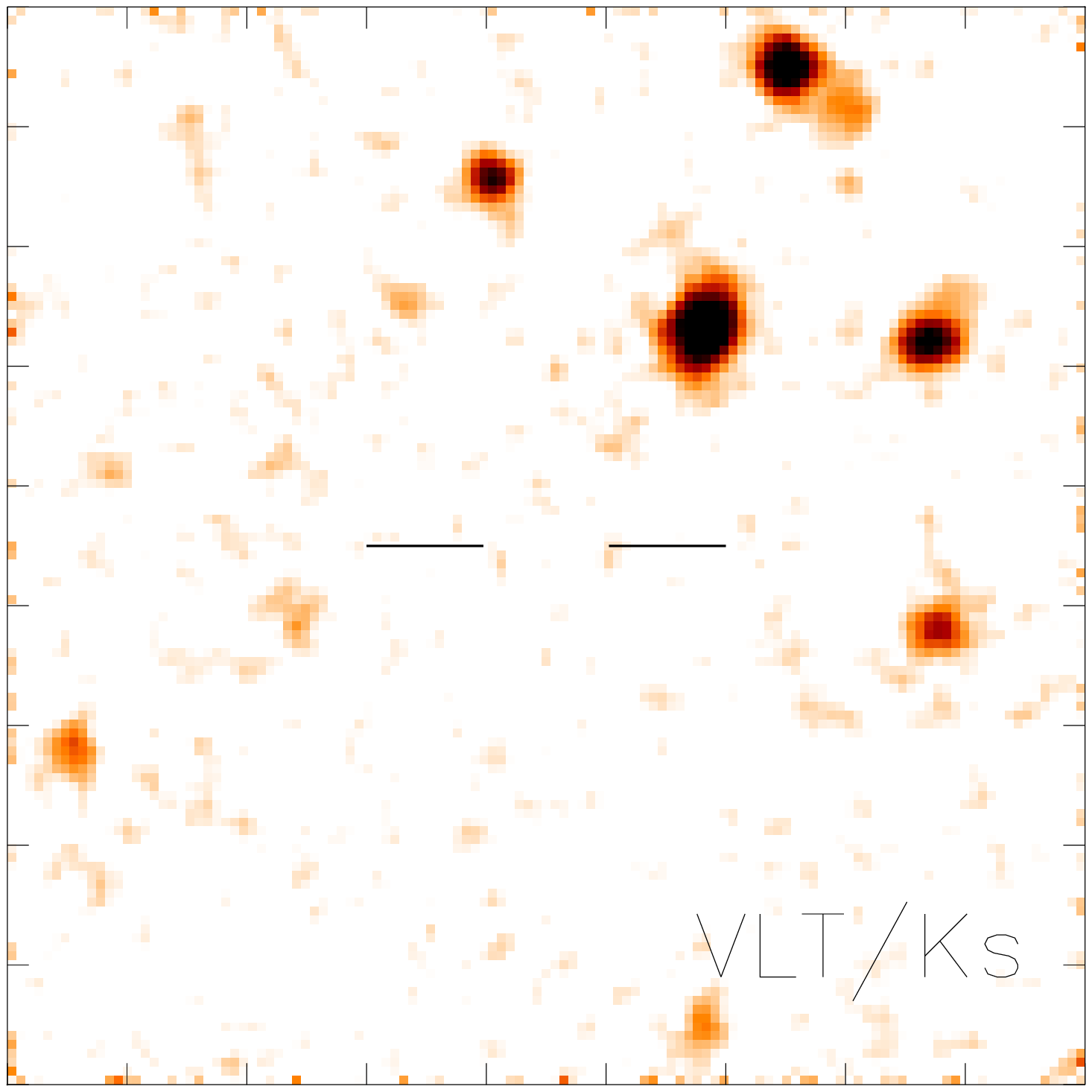,width=2.5cm,clip=}\epsfig{file=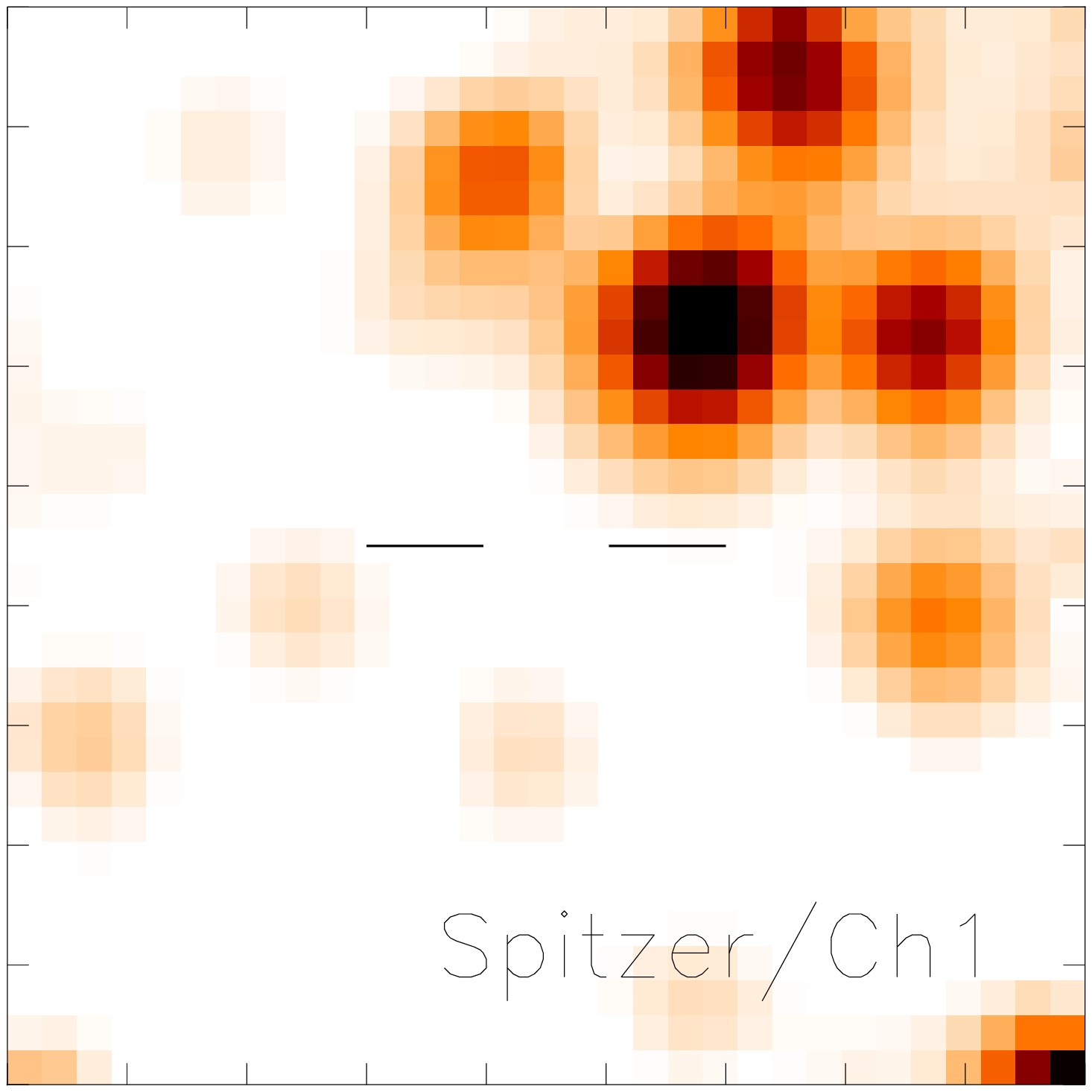,width=2.5cm,clip=}\epsfig{file=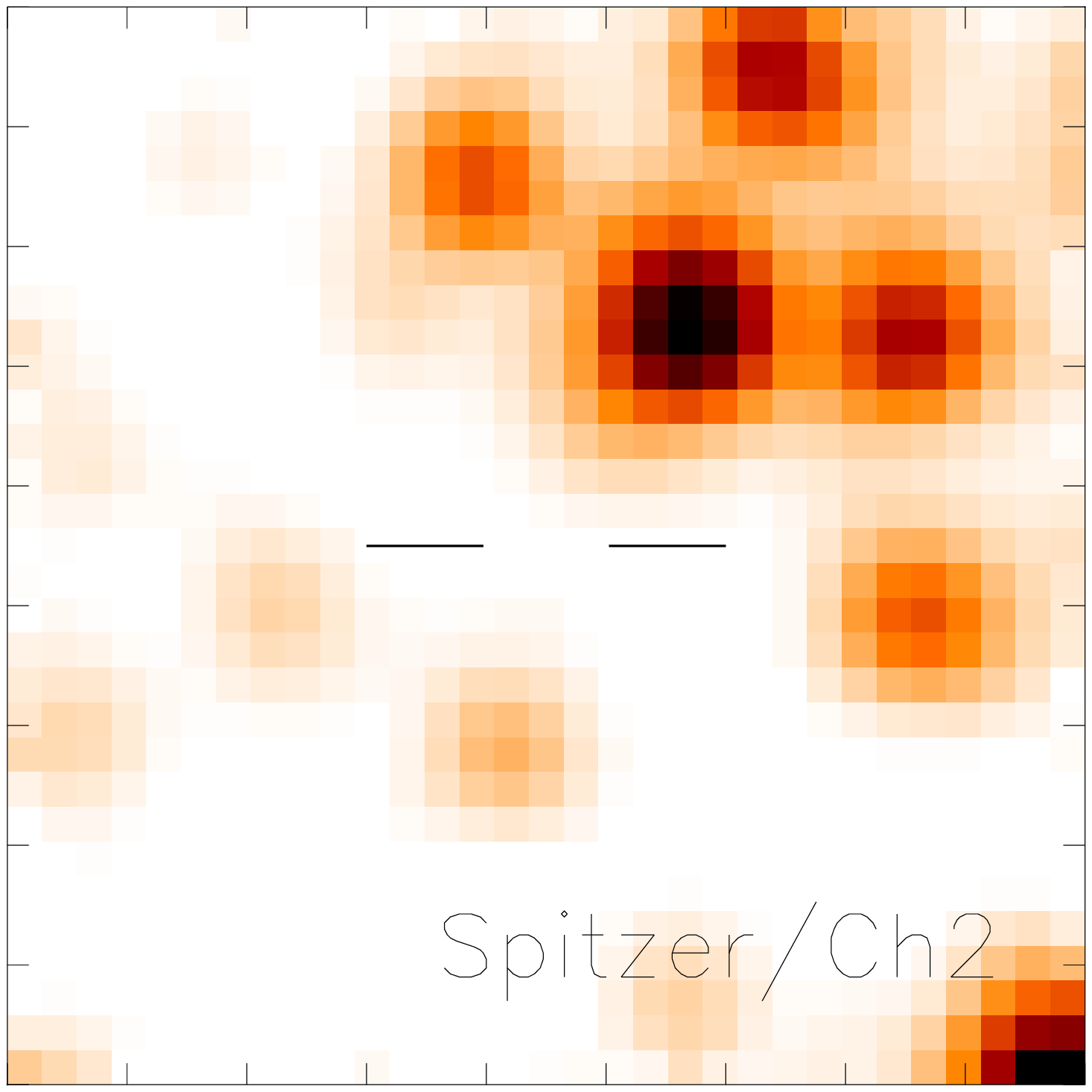,width=2.5cm,clip=}\epsfig{file=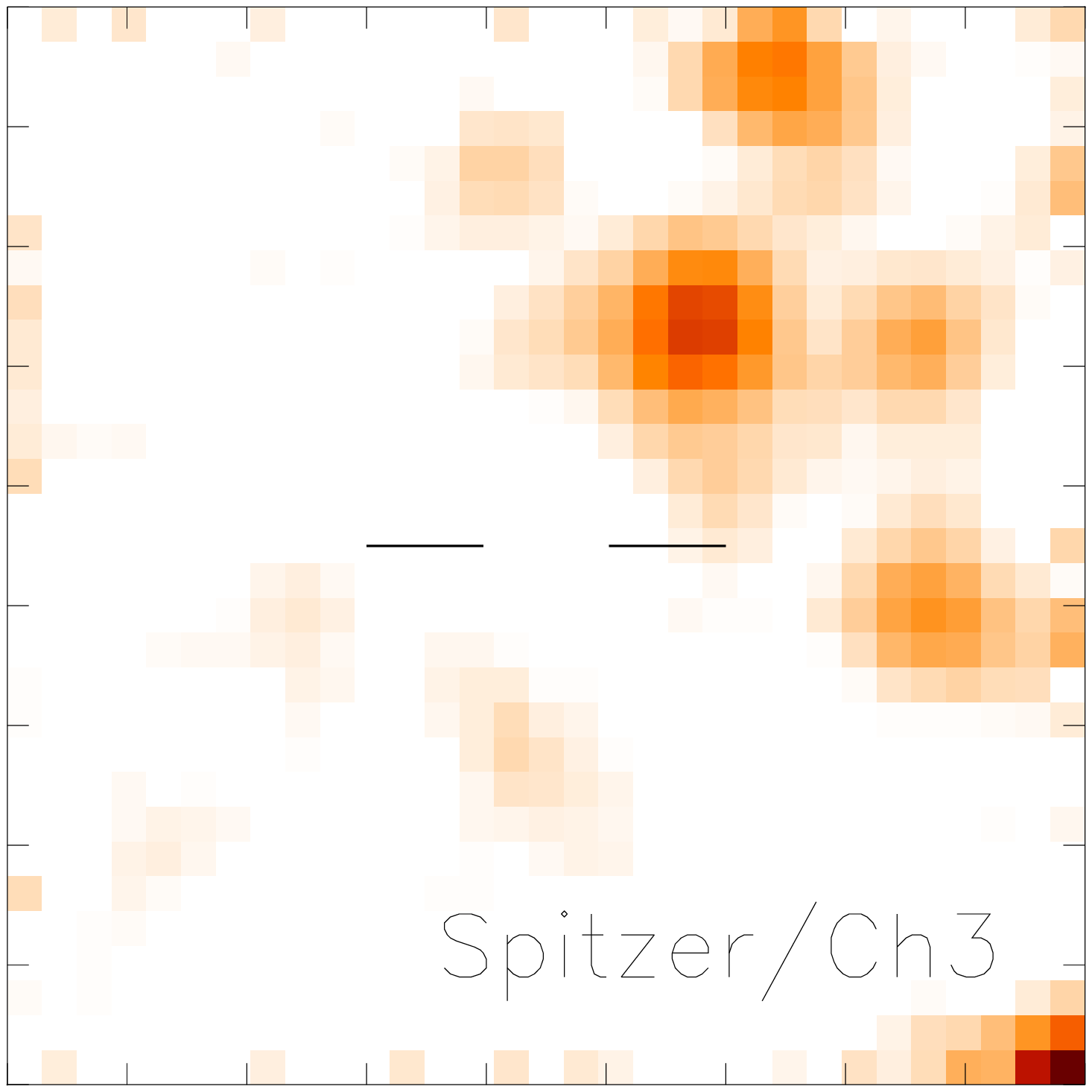,width=2.5cm,clip=}\epsfig{file=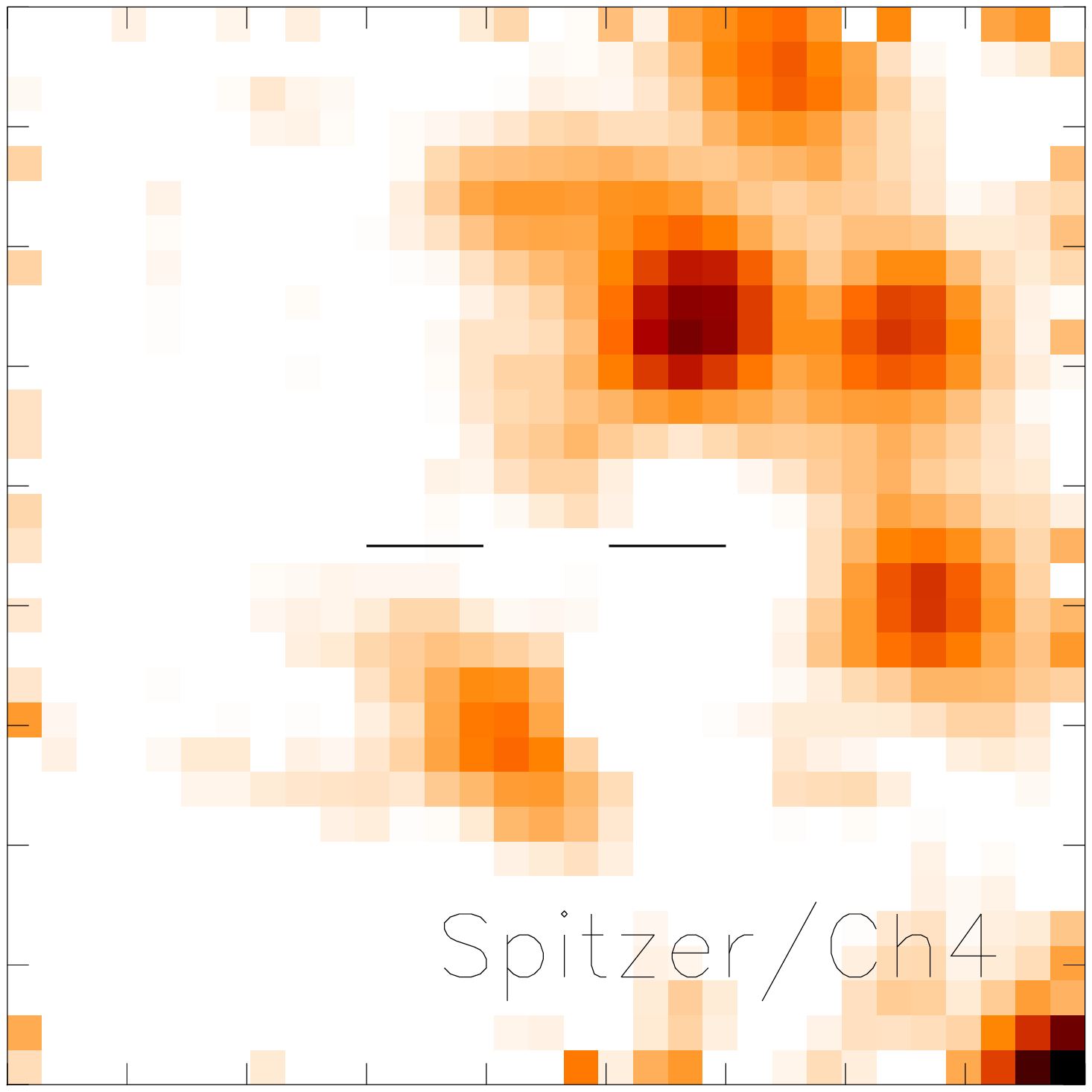,width=2.5cm,clip=}

\caption{Thumbnail images of all available multi-wavelength data in the GOODS South field, centred on the Ly$\alpha$ blob. All images are $18'' \times 18''$.}
\label{thumbs} 
\end{center} 
\vspace{-0.7cm}
\end{figure*}

A contour-plot of the blob superimposed on the HST/ACS
V-band image is shown in the middle panel of Fig.~\ref{contour} and a plot of 
the surface brightness of the blob is seen in the right panel of the same 
figure. The full set of thumb-nail images of the blob 
in all 14 bands can be found in Fig.~\ref{thumbs}. No obvious continuum 
counterpart is detected in any band. The radial 
size is at least 30 kpc (60 kpc diameter) with fainter emission
extending out to 40 kpc radius. This can be seen as the extension to the SW 
in the contour-plot in Fig.~\ref{contour}. The significance of the
lowest contour levels is of the order of $2\sigma$ per pixel. 
The total Ly$\alpha$ luminosity, in a 30 kpc radius aperture, is 
L$_{\mathrm{Ly}\alpha} = (1.09 \pm 0.07) \cdot 10^{43}$~erg~s$^{-1}$. 
This coincides, after correction for the smaller area sampled in the 
spectrum, to the Ly$\alpha$ flux detected in the spectrum within errors. A 
conservative lower limit to 
the restframe equivalent width (EW) of the emission line can be calculated from 
upper limits on the broad-band fluxes in the HST B and V filters in the same
aperture.
This limit is EW~$\gtrsim 220$~{\AA} in the restframe. 
This is in the range of previously published Ly$\alpha$
blobs, that have a Ly$\alpha$
flux to B-band flux density range between 50~--~1500 {\AA} in the restframe
(but typically these values are derived measuring the continuum flux in a 
smaller aperture than the emission line flux).

There are seven objects detected in a wide range ($\ge 8$) of energy bands, 
within a $10''$~radius 
surrounding the blob. 8 other objects are detected within 
the V-band and one further detected in the \emph{Spitzer/IRAC} 
bands. The photometric redshift of these objects was calculated using 
the public \emph{Hyperz}\footnote{http://webast.ast.obs-mip.fr/hyperz/} 
code by Bolzonella et al. (2000). The resulting photometric redshifts for the 
eight objects with most data points ($\ge 8$)
can be found in Table~\ref{photoz}. The other eight objects detected in the
V-band have only a few detections across the spectrum and hence their 
photometric redshifts are unreliable. 
%The galaxy to the west of the 
%blob (object \#~3) 
%was previously given a photometric redshift $z_{phot} = 0.4$ by the 
%COMBO--17 survey (Wolf et al. 2004), but our determination is based on 
%photometry in a 
%larger span of bands. 
The redshift of object \#~3 is similar to the blob redshift and indicates 
that this galaxy may be near to the blob. Object \#~6 is 
an intriguing object, undetected in the deep optical and near-IR imaging 
but bright in the \emph{Spitzer/IRAC} bands. Its photometric redshift is 
consistent with the redshift of the blob, but with a large uncertainty. 
%We investigated the 
%probability to find such an object within a $4''$~radius (corresponding to
%$\lesssim 30$~kpc physical distance, if placed at $z = 3.16$) of the blob by 
%finding all object with similar SEDs in the \emph{Spitzer/IRAC} bands.
%We found the probability to be $\sim 0.3$~\%, hence this object may also be
%associated with the blob.
The object is also detected in the \emph{Spitzer/MIPS} 24~$\mu$m band. 
Based on the Spitzer magnitudes, and on the diagnostic colour-colour
diagram of Ivison et al. (2004), object \#~6 is best fit by a star-burst
at high redshift ($z \sim 5.5$, consistent with the photometric redshift
estimate), hence unrelated to the blob.

\begin{table} 
\begin{center}
\caption{Photometric redshifts of objects surrounding the blob. Numbering 
refers to those given in Fig.~\ref{contour}. Errors given are 1~$\sigma$.
}
\vspace{-0.5cm}
\begin{tabular}{@{}lcccccc}
\hline
Obj \# & Dist. from blob & $z_{phot}$ & $\chi^2/d.o.f$ & Type & A$_V$ rest \\
& (arcsec) & & & & \\
\hline
1   & 4.6 & 1.1$^{+0.40}_{-0.30}$    &  1.3   & Burst  & 0.20 \\
2   & 4.6 & 1.1$^{+0.34}_{-0.41}$    &  8.6   & Burst  & 1.20 \\
3   & 6.8 & 2.9$^{+1.41}_{-0.59}$    &  4.8   & Spiral & 1.20 \\
4   & 8.4 & 0.6$^{+1.97}_{-0.63}$    &  8.1   & Burst  & 0.40 \\
5   & 8.7 & 0.9$^{+2.46}_{-0.89}$    &  2.0   & Burst  & 0.20 \\
6   & 3.0 & 4.5$^{+4.29}_{-1.54}$    &  0.9   & Spiral & 1.20 \\
7   & 6.3 & 1.1$^{+1.24}_{-0.81}$    &  1.9   & Burst  & 1.20 \\
8   & 4.5 & 3.5$^{+1.27}_{-3.48}$    &  0.6   & Spiral & 0.00 \\
\hline
\label{photoz} 
\end{tabular} 
\end{center} 
\vspace{-1.2cm}
\end{table}

\section{Discussion}
We first consider that the Ly$\alpha$ emission of the blob may be due to 
recombination of gas, photo-ionized by an AGN or a starburst galaxy. 
%For a 
%blob of L$_{Ly\alpha}$$\sim$$10^{43}$ erg~s$^{-1}$
%one would, from the (albeit simple) models of Haiman \& Rees
%(2001), expect a characteristic Ly$\alpha$ surface brightness of 
%$\sim10^{-4}$~erg~s$^{-1}$~cm$^{-2}$, significantly less than the peak, as well as typical, surface
%brightness of the blob. 
If the blob is a ``passive'' gas cloud
illuminated and photo-ionized by a nearby AGN, then, following Cantalupo
et al. (2005), one can show that for an AGN with luminosity L$_\nu=
\mathrm{L}_{\rm LL}(\nu/\nu_{\rm LL})^{-\alpha}$, and in order to result 
in a
peak blob Ly$\alpha$ surface brightness of $\Sigma_{Ly\alpha}$, the AGN has to 
be located at a distance of no more than
\[
 270\, \mathrm{kpc} \left(\frac{\Sigma_{Ly\alpha}}{10^{-3}\, \mathrm{erg}\,
\mathrm{s}^{-1} 
\mathrm{cm}^{-2}}\right)^{-1} 
\sqrt{\frac{L_{\rm LL}}{10^{30}\, \mathrm{erg}\, \mathrm{s}^{-1}
\mathrm{Hz}^{-1}}}
\sqrt{\frac{0.7}{\alpha}},
\]
where equality applies to the case where the blob gas is optically
thick at the Lyman limit. No AGN has been detected in the deep \emph{Chandra} 
image available 
within this distance. Worsley et al. (2005) argue that a significant 
fraction 
of the unresolved X-ray background in hard bands consists of highly obscured 
AGN. However, Worsley et al. (2005) also predict 
that the AGN responsible for this background are situated at $z \sim 0.5-1.0$.
Furthermore, Silverman et al. (2005) present a luminosity function for AGN
at higher redshift. To the detection limit of the CDFS (L$_X (z = 3.15) \approx
1.9 \cdot 10^{43}$~erg~s$^{-1}$) and with our search volume 
($3 \times 3$~Mpc~$\times \Delta z = 0.05$) we expect to detect only 0.06 AGN
in our entire search volume. 
We also consider the possibility that galaxy
\#~3 can photo-ionise the blob. However, if we assume a power law for the 
spectrum and extrapolating from the HST/B and HST/V detections we find that 
the UV luminosity of galaxy \#~3 is not sufficient to photo-ionise the blob, 
unless highly collimated towards the blob. We have no reason to believe that 
this is the case.

The second possibility is that the blob Ly$\alpha$ emission
is somehow related to starburst driven, superwind outflows. A starburst would 
be expected to be located within the blob to create such a 
Ly$\alpha$ 
halo and no central continuum source has been detected. Even though a 
very massive starburst can be made invisible in the 
UV/optical range by dust obscuration, it should be visible in the IR, i.e. the 
\emph{Spitzer/IRAC} bands. 

%Moreover, the characteristic bubble
%Ly$\alpha$ emission morphology expected in this case (e.g., Matsuda et al. 2004;
%Mori et al. 2004) is not observed.

The third option is that the Ly$\alpha$ emission is due to cold accretion of
predominantly neutral, filamentary gas onto a massive dark matter halo. 
For cold accretion, the bulk of the Ly$\alpha$ emission is produced 
by collisional
excitation, rather than recombination. Recently, Dijkstra et al. 
2006(a,b) presented a theoretical model for Ly$\alpha$ cooling flows, along 
with predictions of the emission line profile and the shape of the surface 
brightness function. The S/N of our spectrum is not high enough to allow a 
comparison of emission line profiles. However, the surface
brightness profile matches well the predictions for a centrally illuminated
, collapsing
cloud of Dijkstra et al. 2006(a), see Fig.~1. Further tests are needed
to determine how well their model fits.
To test whether this blob can be filamentary gas accreting
``cold'' onto a companion galaxy, we also conducted the following 
experiment:
we calculated the Ly$\alpha$ surface brightness
in a 100$\times$100 kpc (projected) region for a proto-galaxy
of ``cooling'' radiation only (so all contributions from regions with young 
stars were removed, as well as all emission, in general, from gas closer
than 10 kpc to any star-forming region). The calculation was based
on a cosmological simulation of the formation and evolution of an
M31-like disk galaxy (Sommer-Larsen 2005; Portinari \& Sommer-Larsen
2005). 

The results at $z\sim3$ are presented in Sommer-Larsen (2005),
and get to
a surface brightness about an order of magnitude lower than the observed level.
This is interesting, and may point to a cold accretion origin
of the blob Ly$\alpha$ emission on a larger scale, such as
filamentary gas accretion onto a galaxy-group sized halo.
Another possibility is that the periods with high surface brightness 
are shorter than 2.5 Myr (the resolution of the simulation). Given that in a 
search volume of about
40000 co-moving Mpc$^3$, only one such blob has been detected, it is
actually comforting, that we could not reproduce the blob
characteristics, by cold accretion onto this, randomly selected, M31-like 
galaxy. This has to be a rare phenomenon.

A test for the cold accretion model would be to observe the Balmer lines.
For collisionally excited hydrogen, neglecting extinction
effects, the flux in H$\alpha$ should only be about 3.5 percent of the
Ly$\alpha$ flux, whereas for recombining, photo-ionized gas this ratio is $\sim
11.5$~\% (Brocklehurst 1971). Hence, the relative H$\alpha$ luminosity is
expected to be significantly larger in the latter case. The situation is
similar for H$\beta$, and whereas the H$\alpha$ line will be very difficult to
detect from the ground, H$\beta$ should be observable.

\section{Conclusion}
We have here reported the results of an extensive multi-wavelength
investigation of a redshift $z = 3.16$ Ly$\alpha$~emitting blob discovered in
the GOODS South field. The blob has a diameter larger than 60~kpc
diameter and a total luminosity of $\mathrm{L}_{\mathrm{Ly}\alpha} \sim
10^{43}$~erg~s$^{-1}$. Deep HST imaging show no obvious optical counterpart,
and the lack of X-ray or IR emission suggest there are no AGN or dusty
starburst components associated with at least the centroid of the blob. 
Two galaxies within a $10''$ radius have photometric redshifts consistent
with the redshift of the blob, but follow-up spectroscopy is needed to
establish if there is a connection.
We have run simulations of Ly$\alpha$ surface brightness arising from
cold accretion and found that such extended Ly$\alpha$ emission may be
explained by accretion 
along a filament onto a galaxy group sized dark matter halo. Another
possibility is that such emission in very short lived, i.e. significantly 
shorter than the 2.5 Myr resolution of our simulation. We argue that other
previously suggested origins of Ly$\alpha$ blobs (hidden AGN and 
``super-winds'')
can be ruled out in this case due to the lack of detected continuum 
counter-parts. Hence, 
though our cold accretion simulation cannot perfectly match our data, it is the
only explanation that is plausible. Our results combined 
with 
the fact that previously studied blobs appear to be caused by superwinds 
and/or AGN in turn implies that the energy sources for blob Ly$\alpha$ 
emission are diverse. 

\vspace{-0.1cm}
\begin{acknowledgements}
KN gratefully acknowledges support from IDA~--~Instrumentcentre for Danish
Astrophysics. The numerical simulations used in this paper were performed on
the facilities provided by Danish Center for Scientific Computing (DCSC). The
Dark Cosmology Centre is funded by the DNRF. The authors thank the DDT panel 
and the ESO Director General for granting time for follow-up spectroscopy. 
\end{acknowledgements}

\end{document}